\documentclass[reprint,pra]{revtex4-2}
\usepackage{amsmath}
\usepackage{amssymb}
\usepackage{physics}
\usepackage{color}
\usepackage{graphicx}
\usepackage{float}
\usepackage{hyperref}
\usepackage{bbold}
\usepackage[utf8]{inputenc}

\begin{document}
\title{Gaussian Continuous-Variable Isotropic State}
\author{Maria Poxleitner}
\email{maria.poxleitner@physik.uni-wuerzburg.de}
\author{Haye Hinrichsen}
\email{haye.hinrichsen@physik.uni-wuerzburg.de}
\affiliation{Faculty for Physics and Astronomy, University of Würzburg, Campus Süd, Am Hubland, 97074 Würzburg, Germany}
\date{\today}

\begin{abstract}  
Inspired by the definition of the non-Gaussian two-parametric continuous variable analogue of an isotropic state introduced by Mi\v{s}ta \textit{et al.} [Phys. Rev. A, \textbf{65}, 062315 (2002)], we propose to take the Gaussian part of this state as an independent state by itself, which yields a simple, but with respect to the correlation structure interesting example of a two-mode Gaussian analogue of an isotropic state. Unlike conventional isotropic states which are defined as a convex combination of a thermal and an entangled density operator, the Gaussian version studied here is defined by a convex combination of the corresponding covariance matrices and can be understood as entangled pure state with additional Gaussian noise controlled by a mixing probability. Using various entanglement criteria and measures, we study the non-classical correlations contained in this state. Unlike the previously studied non-Gaussian two-parametric isotropic state, the Gaussian state considered here features a finite threshold in the parameter space where entanglement sets in. In particular, it turns out that it exhibits an analogous phenomenology as the finite-dimensional two-qubit isotropic state.
\end{abstract}

\keywords{Isotropic states, Werner states, Gaussian states, entanglement, quantum discord}
\maketitle

\section{Introduction}
\label{sec:introduction}

Entanglement is one of the most intriguing phenomena in quantum physics and serves as a resource for many applications in quantum information processing, particularly in quantum communications~\cite{horodeckis1}. A bipartite system is defined to be entangled if its joint quantum state is not separable. It is well known and experimentally confirmed that entangled systems exhibit quantum correlations which cannot be explained in solely classical terms. What is less known is that the opposite implication does not apply. In fact, it turned out that also separable, i.e. non-entangled states, can exhibit non-classical correlations which are, for instance, relevant for quantum computational tasks~\cite{datta1,lanyon,ferraro}. 

In order to understand the difference between the two types of quantum correlations, it is useful to study simple bipartite quantum systems which cross over from a non-classical separable to an entangled state when a parameter is varied. Examples of this type include so-called \textit{Werner states}~\cite{Werner1} and \textit{isotropic states}~\cite{horodeckis2}. 

An isotropic state is defined as a convex combination of a totally mixed and a fully entangled state. More specifically, for a bipartite system consisting of two $d$-dimensional subsystems $A$ and $B$, the isotropic state is defined by
\begin{equation}
\hat{\rho}_\text{{\tiny I}}\,=\,p\dyad{+}\,+\,(1-p)\frac{\mathbb{1}}{d^2}
\label{eq:iso_state}
\end{equation}
with a mixing parameter $p\,\in\,[0,1]$, where
\begin{equation}
\ket{+}\,=\,\frac{1}{\sqrt{d}}\sum\limits_{i=1}^d\ket{i,i}\,=\,\frac{1}{\sqrt{d}}\sum\limits_{i=1}^d\ket{i}_A\otimes\ket{i}_B.
\end{equation}
This state depends on a single parameter $p\in [0,1]$ which determines the balance between the mixed and the entangled component. Varying this parameter, it turns out that the isotropic state is separable below a certain finite threshold $p_c=\frac{1}{1+d}$ and entangled above~\cite{horodeckis2}. In the range $0 < p \leq p_c$ the isotropic state offers an interesting example of a quantum state with zero entanglement which nevertheless exhibits quantum correlations that cannot be explained in classical terms, as will be discussed below. 

Werner states are closely related to isotropic states. While isotropic states are defined to be invariant under local unitaries of the form $U \otimes U^*$, Werner states are invariant under $U \otimes U$. Werner states in finite-dimensional systems are known to exhibit similar properties as isotropic ones, in particular there is also a finite threshold for the control parameter from where on entanglement sets in. In the two-qubit case, Werner and isotropic states are identical up to a local unitary transformation~\cite{horodeckis2}. 

The most common measure for quantifying quantum correlations in mixed states is the so-called \textit{entanglement of formation} (EOF)~\cite{bennet,wotters}. Here, the statistical ensemble described by the density matrix is decomposed into pure-state components and the corresponding entanglement entropy is then averaged according to the statistical weights. Since a mixed quantum state represents an equivalence class of many statistical ensembles, the result has to be minimized over all possible ensembles in this class, which is a technically difficult task. 

Since the EOF vanishes on separable states, it is not suitable for quantifying the aforementioned quantum correlations in the separable region of the isotropic state. Contrarily, the so-called quantum discord (QD) is a measure that does also respond to quantum correlations in the separable region. The QD is defined as the difference between total and classical correlations. Whereas the former are measured by the quantum mutual information, the latter are extractable via quantum measurement, where one has to perform an optimization over all possible quantum measurements~\cite{ollivier, henderson}. In fact, for two-qubit isotropic states, the QD was found to be non-zero for all $0<p\leq 1$~\cite{ollivier}.

In quantum information theory, besides systems with finite-dimensional Hilbert spaces, infinite-dimensional systems with continuous variables (CV) are studied frequently~\cite{braunstein,adesso2014}. Experimentally such systems arise naturally in quantum optics where quantum states can be realized by light modes. In this context, Gaussian states play a central role since they are fully specified by a displacement vector and a finite-dimensional covariance matrix~\cite{adesso, weedbrook}. From an experimental point of view, Gaussian states are of great importance because they can be created with simple optical tools -- the generation of squeezed states using optical parametric oscillators~\cite{Wu_87} or deterministic entanglement creation using optical parametric amplifiers~\cite{Zhang_2000} are well known. Moreover, entangled Gaussian states are fundamental resources for many CV quantum communication protocols such as teleportation~\cite{Furusawa_1998,Huo_2018} or key distribution~\cite{Grosshans_2002,Garcia_Patron_2009,Wang_2018,Wang_19}. In addition, the Gaussian CV setting is ideally suited for the generation of large-scale entangled states -- so-called cluster states -- which are highly relevant for measurement-based quantum computation~\cite{Chen_2014,Asavanant_2019,Larsen_2019}.

Given the importance of CV quantum systems both in theory and experiment, the question arises whether it is possible to define a continuous variant of isotropic and Werner states with similar properties. The first attempt in this direction was made by Mi\v{s}ta~\textit{et al.}, who introduced a two-parametric CV analogue as a convex combination of a two-mode thermal and a two-mode squeezed state~\cite{mista}. However, this CV isotropic state is generally non-Gaussian and therefore the well-established formalism for Gaussian states cannot be applied. Furthermore, it turns out that the suggested two-parametric state does not feature a finite entanglement threshold, instead it is entangled for all $p>0$.

In this paper, we propose to take the purely Gaussian part of the non-Gaussian CV isotropic state and to consider it as an \textit{independent} state for its own right. In contrast to the non-Gaussian state introduced by Mi\v{s}ta~\textit{et al.}, which was defined as a convex combination of two density matrices, the state proposed here is defined as a convex combination of the covariance matrices and hence it is Gaussian by construction. The analysis of this state is therefore relatively simple since the well-known results for Gaussian states can be directly applied. Unlike the previously studied state, the Gaussian version does exhibit a finite threshold in the parameter space where entanglement sets in. Furthermore, it turns out that it displays analogies to the finite-dimensional two-qubit isotropic state. The aim of the present work is to study the correlation structure of this Gaussian CV isotropic state and to understand the quantum nature of the correlations in the separable domain.

The paper is organized as follows. In the next section, we briefly review the common notations of CV systems. Sect.~\ref{sec:cv_iso_state} reviews the non-Gaussian isotropic state introduced by Mi\v{s}ta~\textit{et al.} while the Gaussian version studied here is defined in Sect.~\ref{sec:gaussian_example}. After analyzing various properties of this state in Sect.~\ref{sec:properties}, various entanglement criteria and measures of quantum correlations are applied in Sect.~\ref{sec:criteria} and \ref{sec:measures}. In Sect.~\ref{sec:gaussian_channel}, we describe the Gaussian channel isomorphic to the Gaussian isotropic state.  After concluding remarks, the corresponding density matrix is calculated explicitly in the Appendix.

\section{Gaussian states}
\label{sec:gaussian_states}

Before starting let us briefly recall some common notations for CV systems. A quantum $N$-mode system is given by $N$~pairs of canonical observables 
\begin{equation}
\hat{x}_k=\frac{1}{\sqrt{2}}\left(\hat{a}^\dagger_k+\hat{a}_k\right),\quad
\hat{p}_k=\frac{i}{\sqrt{2}}\left(\hat{a}^\dagger_k-\hat{a}_k\right)
\label{eq:canon_obs}
\end{equation}
with $[\hat{a}_k,\hat{a}_l^\dagger]=\delta_{kl}$. Arranging them in a vector
\begin{equation}
\vec{R}=\left(\hat{x}_1,\hat{p}_1,...,\hat{x}_N,\hat{p}_N\right)=\left(R_1,...,R_{2N}\right)
\end{equation}
they obey the commutation relations $[R_k, R_l]\,=\,i\Omega_{kl}\mathbb{1}$, where
\begin{equation}
\Omega\,=\,\bigoplus\limits_{k=1}^N\left(\begin{array}{cc}0&1\\-1&0\end{array}\right)
\end{equation}
denotes the symplectic matrix. Any quantum state $\hat\rho$ of such a CV system can be expressed as an integral over all points $\vec r=(r_1,...,r_{2N})=({x}_1,{p}_1,...,{x}_N,{p}_N)$ in phase space by
\begin{equation}
	\hat{\rho}\,=\,\frac{1}{(2\pi)^N}\int_{-\infty}^\infty d^{2N}r\,\,\chi(\vec{r})\,\hat{D}^\dagger(\vec{r})\,,
\label{eq:integral_relation}
\end{equation}
where $\hat{D}(\vec{r})=e^{-i\vec{r}^T\Omega\vec{R}}$ is the displacement operator and $\chi(\vec r)=\text{Tr}[\hat\rho \hat D(\vec r)]$ is the characteristic function. 

A quantum state $\hat\rho$ is called \textit{Gaussian} if its characteristic function can be written in the quadratic form
\begin{equation}
	\chi_{\hat{\rho}}(\vec{r})\,=\,\exp(i\vec{r}^T\Omega\vec{d}-\frac{1}{4}\vec{r}^T\Omega^T\gamma\Omega\vec{r})\,.
\label{eq:CF}
\end{equation}
This means that a Gaussian state is completely determined by its first two cumulants, namely, the displacement $\vec d=\langle \vec R \rangle$ and the covariance matrix (CM) $\gamma$ with the components $\gamma_{kl}\,=\,\langle R_kR_l+R_lR_k\rangle-2\langle R_k\rangle\langle R_l\rangle$. 

The uncertainty relation manifests itself in the fact that all valid quantum states fulfill the condition~\cite{simon1}
\begin{equation}
	\gamma\,+\,i\Omega\,\geq\,0
\label{eq:phys_cond}
\end{equation}
in the sense that the matrix sum on the l.h.s. has only non-negative eigenvalues. For Gaussian states, this condition is both necessary and sufficient for physicality.

When discussing correlation properties of Gaussian states, the displacement $\vec d$ is usually not relevant, meaning that the CM $\gamma$ alone characterizes the correlation structure of the state. Moreover, for two-mode Gaussian states we can use local symplectic transformations to convert a given CM into the following standard form~\cite{simon2,duan}
\begin{equation}
	\gamma\,=\,\left(\begin{array}{cccc}
												a & 0 & c_1 & 0\\
												0 & a & 0 & c_2\\
												c_1 & 0 & b & 0\\
												0 & c_2 & 0 & b\\
												\end{array}\right),
\quad a,b,c_1,c_2 \in \mathbb{R}.												
\label{eq:standard_form}
\end{equation}
If $a\,=\,b$, the two-mode Gaussian state is said to be \textit{symmetric}. In addition, according to the Williamson theorem~\cite{williamson1936}, a general $N$-mode CM can always be brought into the so-called canonical or normal form
\begin{equation}
\gamma_n\,=\,\text{diag}\bigl(\nu_1,\nu_1,...,\nu_{N},\nu_{N}\bigr)
\label{eq:normal_form}
\end{equation}
by global symplectic transformations, where the $\nu_k\geq 1$ are the symplectic eigenvalues which can be obtained as eigenvalues of $|i\Omega\gamma|$ (see \cite{adesso} and references therein). In case of two-mode Gaussian states, these symplectic eigenvalues are functions of the entries in Eq.\,(\ref{eq:standard_form}). For two-mode Gaussian states, all correlation properties, such as entanglement and Gaussian quantum discord, can be fully discussed in terms of these matrix entries or the symplectic eigenvalues~\cite{adesso, datta}.

\section{Non-Gaussian CV isotropic state}
\label{sec:cv_iso_state}

Let us briefly review the CV isotropic state introduced by Mi\v{s}ta \textit{et al.} in Ref.~\cite{mista}. In order to define a CV analogue of the isotropic state in Eq.~(\ref{eq:iso_state}), they proposed to consider a convex combination of a two-mode squeezed and a two-mode thermal state
\begin{equation}
	\hat{\rho}^\text{{\tiny $\infty$}}_\text{{\tiny I}}\,:=\,p\hat{\rho}_\text{{\tiny TMS}}\,+\,(1-p)\hat{\rho}_\text{{\tiny TMT}},\quad p\,\in\,[0,1]\,.
\label{eq:CV_iso_state}
\end{equation}
Here the entangled component $\hat{\rho}_\text{{\tiny TMS}}$ is a pure two-mode squeezed (TMS) vacuum state~\cite{bruss2} which can be represented in the Fock basis $\ket{m,n}\,=\,\ket{m}_A\otimes\ket{n}_B$ of the two modes $A$ and $B$ by
\begin{equation}
	\hat{\rho}_\text{{\tiny TMS}}\,=\,(1-\lambda_1)\sum\limits_{m,n=0}^{\infty}\lambda_1^{\frac{m+n}{2}}\dyad{m,m}{n,n}\,,
\label{eq:TMS}
\end{equation}
where $\lambda_1\,=\,\tanh^2r$ depends on the so-called squeezing parameter $r \in \mathbb{R}$. The physical meaning of this parameter is that it controls the average particle number $\bar n=\sinh^2 r$. It is important to note that $\hat{\rho}_\text{{\tiny TMS}}$ itself is only \textit{partially} entangled and becomes fully entangled only in the limit $r\to\infty$, but this limit is unphysical since then the particle number diverges and the state is no longer normalizable. For this reason, one has to keep $r$ as a free parameter.

The separable mixed component $\hat{\rho}_\text{{\tiny TMT}}$ is a two-mode thermal (TMT) state, given as product of two identical thermal modes 
\begin{equation}
	\hat{\rho}_\text{{\tiny TMT}}\,=\,(1-\lambda_2)^2\sum\limits_{m,n=0}^{\infty}\lambda_2^{m+n}\dyad{m,n},
\label{eq:TMT}
\end{equation}
where $\lambda_2\,=\,\tanh^2s$ is another parameter controlling the average particle number $\bar{n}\,=\,\sinh^2s$. 

Thus, the proposed state in Eq.~(\ref{eq:CV_iso_state}) depends on three parameters, $r$, $s$, and $p$. However, in order to establish a close analogy with the finite-dimensional case, it is sufficient to restrict the analysis to the case $r=s$ since then both components, the TMT and the TMS state, involve the same average number of particles. Moreover, this choice ensures that the TMT state is just twice the reduced TMS state, i.e. 
\begin{equation}
\hat{\rho}_\text{{\tiny TMT}}\,=\,\Tr_B\left[\hat{\rho}_\text{{\tiny TMS}}\right]\otimes\Tr_A\left[\hat{\rho}_\text{{\tiny TMS}}\right]\,,
\end{equation}
analogous to Eq.~(\ref{eq:iso_state}) where a similar relation holds. With this restriction, the state is controlled by only two parameters $r$ and $p$. Note that in the following, we will always refer to this \textit{two}-parametric version of the CV isotropic state.

In Ref.~\cite{mista}, Mi\v{s}ta \textit{et al.} showed that the two-parametric CV isotropic state exhibits two important properties. Firstly, it is \textit{non-Gaussian} for all $0<p<1$ so that it can no longer be described in terms of the CM matrix alone. Secondly, the state is \textit{entangled} (non-separable) for all $p>0$ irrespective of $r$. At a first glance this is surprising since in the finite-dimensional case one observes a finite threshold $p>p_c=\frac{1}{1+d}$ from where on entanglement sets in. Yet the result is consistent if we recall that a CV system involves infinitely many degrees of freedom, which would correspond to taking the limit $d\to\infty$~\cite{mista}.

However, we should keep in mind that a direct analogy between the finite-dimensional and the CV case is only valid in the limit $r\,\to\,\infty$ where $\hat{\rho}_\text{{\tiny TMS}}$ becomes maximally entangled and $\hat{\rho}_\text{{\tiny TMT}}$ becomes maximally mixed. Nevertheless, $\hat{\rho}^\text{{\tiny $\infty$}}_\text{{\tiny I}}$ is entangled for all $p>0$ \textit{independent} of the squeezing parameter $r$. This means that in contrast to the finite-dimensional isotropic state, the two-parametric CV state introduced by Mi\v{s}ta \textit{et al.} does not feature an extended region in the parameter space where the state is non-entangled but nevertheless non-classical. If one is primarily interested in understanding the 'quantumness' of such non-entangled states and the transition into the entangled regime, the CV isotropic state defined in Eq.~(\ref{eq:CV_iso_state}) is perhaps less interesting as the finite-dimensional one. 

To conclude this section, let us remark that further properties of the non-Gaussian CV isotropic state, including non-classical correlations beyond entanglement, have been studied extensively in Refs.~\cite{tatham_2012,McNulty_2014}. Furthermore, a special variant of the state proposed by Mi\v{s}ta \textit{et al.} is investigated in Ref.~\cite{Lund_2006} which aims at elaborating possible advantages of non-Gaussian states over their ``closest'' Gaussian version.

\section{Gaussian isotropic state}
\label{sec:gaussian_example}

Besides being of great interest in non-Gaussian CV quantum information, the properties of the state introduced in Ref.~\cite{mista} -- non-Gaussianity and entanglement in the whole parameter space -- has to be considered as a drawback if one is looking for a simple CV system to study non-classical correlations in the absence of entanglement. Here, we would prefer a two-mode Gaussian state since these are the simplest representatives of bipartite CV quantum systems. Therefore, we consider a different kind of interpolation between $\hat{\rho}_\text{{\tiny TMT}}$ and $\hat{\rho}_\text{{\tiny TMS}}$. The idea is very simple: instead of taking a convex combination of the density matrices, we consider a convex combination of the corresponding covariance matrices
\begin{equation}
	\gamma_\text{{\tiny GI}}\,:=\,p\gamma_\text{{\tiny TMS}}\,+\,(1-p)\gamma_\text{{\tiny TMT}}
\label{eq:gamma_GI1}
\end{equation}
and use this combination to define a Gaussian state $\hat\rho_\text{{\tiny GI}}$ via Eqs. (\ref{eq:integral_relation})-(\ref{eq:CF}) that interpolates between a thermal and an entangled squeezed state. Like $\hat{\rho}^\text{{\tiny $\infty$}}_\text{{\tiny I}}$, this state depends on two parameters, namely, the mixing parameter $p\in[0,1]$ and the squeezing parameter $r\in \mathbb{R}$.

To construct the CM $\gamma_\text{{\tiny GI}}$, we note that $\gamma_\text{{\tiny TMS}}$ is already given in the standard form (\ref{eq:standard_form}) with $a=b=\cosh(2r)$ and $c_1=-c_2=\sinh(2r)$~\cite{adesso,bruss2}. Concerning $\gamma_\text{{\tiny TMT}}$, it is known that an $N$-mode thermal state has a diagonal CM of the form (\ref{eq:normal_form}), where the symplectic eigenvalues are related to the average number of mode-excitations via $\nu_k\,=\,2\bar{n}_k+1$~\cite{bruss1}. Since we consider two identical thermal states which are the reduced states of the TMS state, we have $\nu_1=\nu_2=1+2\sinh^2r=\cosh(2r)$. Combining all contributions, we obtain the CM
\begin{equation}
	\gamma_\text{{\tiny GI}}\,=\,\left(\begin{array}{cccc}
        \cosh 2r & 0 & p\sinh 2r & 0\\
        0 & \cosh 2r & 0 & -p\sinh 2r\\
        p\sinh 2r & 0 & \cosh 2r & 0\\
        0 & -p\sinh 2r & 0 & \cosh 2r\\
        \end{array}\right)
\label{eq:gamma_GI2}
\end{equation}
which in turn determines the full quantum state via
\begin{equation}
\hat\rho_\text{{\tiny GI}}\;=\;
\frac{1}{4 \pi^2}\int \text{d}^4 r \, e^{-\frac14 (\Omega \vec r)^T \gamma_\text{{\tiny GI}}\,\, \Omega \vec r}
D^\dagger(\vec r)\,.
\label{densitymatrixGI}
\end{equation}
It can be checked easily that $\gamma_\text{{\tiny GI}}$ fulfills the condition~(\ref{eq:phys_cond}), confirming that it represents a physically valid quantum state. An explicit matrix representation of the density operator $\hat\rho_\text{{\tiny GI}}$ is given in Appendix~\ref{sec:density}. 

What is the difference between $\hat{\rho}^\text{{\tiny $\infty$}}_\text{{\tiny I}}$ and $\hat\rho_\text{{\tiny GI}}$? First, we note that the CM  of the previously studied state $\hat{\rho}^\text{{\tiny $\infty$}}_\text{{\tiny I}}$ is given by exactly the same expression as in Eq.~(\ref{eq:gamma_GI2})~\cite{mista}, that is, both states are characterized by the same CM. However, for $0<p<1$ the mixed state defined in Eq.~(\ref{eq:CV_iso_state}) is non-Gaussian, meaning that it involves also unknown higher cumulants, while the state~$\hat\rho_\text{{\tiny GI}}$ is Gaussian by construction and -- up to possible local displacements -- completely determined by $\gamma_\text{{\tiny GI}}$ alone. At this point, however, we should emphasize that it is in principle possible to obtain $\hat\rho_\text{{\tiny GI}}$ from $n\,\to\,\infty$ many copies of $\hat{\rho}^\text{{\tiny $\infty$}}_\text{{\tiny I}}$ by a Gaussification operation based on the quantum central limit theorem as discussed in Ref.~\cite{wolf_2006},\cite{mark_wilde}. In fact, $\hat\rho_\text{{\tiny GI}}$ is simply the purely Gaussian part of $\hat{\rho}^\text{{\tiny $\infty$}}_\text{{\tiny I}}$, but here, we consider it as an independent state which is interesting for its own right as will turn out below. Note that a special variant of $\gamma_\text{{\tiny GI}}$, where the thermal part is replaced by a two-mode vacuum, is mentioned in Ref.~\cite{Lund_2006} but investigated within another context.

\begin{figure}
\begin{center}
\includegraphics[width=65mm]{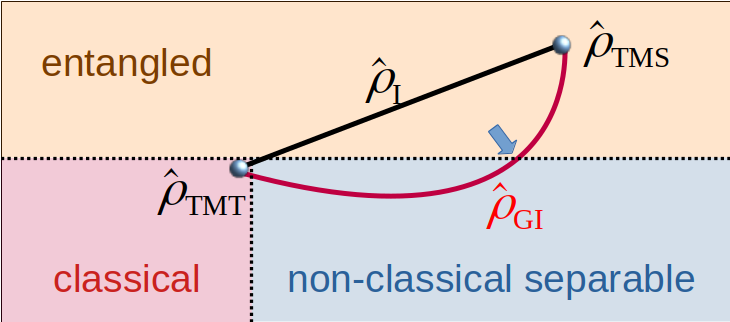}                           \caption{Cartoon of the space of quantum states, illustrating the difference between $\hat{\rho}^\text{{\tiny $\infty$}}_\text{{\tiny I}}$ and $\hat\rho_\text{{\tiny GI}}$ (see text).}
\label{fig:cartoon}
\end{center}
\end{figure}

Fig.~\ref{fig:cartoon} illustrates the conceptual difference between $\hat\rho_\text{{\tiny GI}}$ and $\hat{\rho}^\text{{\tiny $\infty$}}_\text{{\tiny I}}$. In this figure, the drawing plane represents a cartoon of the space of all density matrices of a two-mode CV system. This space comprises three subspaces, containing separable classical, separable non-classical, and entangled states, respectively. Both definitions, $\hat{\rho}^\text{{\tiny $\infty$}}_\text{{\tiny I}}$ and $\hat\rho_\text{{\tiny GI}}$, interpolate between the two-mode thermal state $\hat{\rho}_\text{{\tiny TMT}}$ and the two-mode squeezed state $\hat{\rho}_\text{{\tiny TMS}}$ when $p\in[0,1]$ is varied. However, they do so along different paths. For the CV isotropic state introduced by Mi\v{s}ta \textit{et al.}, which is defined as a convex combination of its ending points, the interpolation path is indicated as a black straight line which lies entirely in the region of entangled states. Contrarily, $\hat\rho_\text{{\tiny GI}}$ connects the two ending points in a different way, sketched here as a curved line. As we will see below, this line crosses over from the separable non-classical domain into the entangled region at a finite threshold of $p$ marked by the arrow. 

Any mixed Gaussian state can be thought as being composed of a pure Gaussian state with additional Gaussian noise, i.e. $\gamma_\text{{\tiny mixed}}\,=\,\gamma_\text{{\tiny pure}}+\gamma_\text{{\tiny noise}}$. In Hilbert space, this can be understood as convolution of a pure density operator with a classical Gaussian probability distribution~\cite{krueger}, which suggests a variety of possibilities to generate such a state in experiments. One of the simplest possibilities would be to transmit each mode of a TMS through identical channels exposed to thermal loss~\cite{serafini2004}, interpreting $\gamma_\text{{\tiny GI}}$ as a pure TMS attenuated with two thermal-loss channels. More generally, the influence of Gaussian noise is an unavoidable side effect in most experimental realizations (see e.g. Refs.~\cite{serafini2005,Garcia_Patron_2009,goyal2010,Wang_2018}), suggesting that the state $\gamma_\text{{\tiny GI}}$ may appear in a wide range of physical situations.

The state (\ref{eq:gamma_GI2}) may also play a role in the context of quantum teleportation. For instance, in the Wigner representation, the output state of the quantum teleportation protocol described in Ref.~\cite{Braunstein_Kimble_1998} is given by the convolution of a pure input state with Gaussian noise. This is also the case for measurement-based quantum computation where teleportation is part of the computation scheme~\cite{Larsen_2020}. However, in order to control the parameter $p$, it would be necessary to generalize these protocols from unity gain to the non-unity gain regime, where the first moments of the teleported state are not preserved.

\section{General properties}
\label{sec:properties}
%
In the following, we summarize some basic properties of the proposed Gaussian isotropic state.
The symplectic eigenvalues of the CM defined in (\ref{eq:gamma_GI2}) are 
\begin{equation}
\nu:=\nu_1=\nu_2\;=\;\sqrt{\cosh^2(2r)-p^2\sinh^2(2r) }\;\geq\; 1,
\end{equation}
so that the purity of the state is given by (see~\cite{serafini})
\begin{equation}
\mu \;=\; \Tr[\hat\rho_\text{{\tiny GI}}^2] \;=\; \frac{1}{\sqrt{\det\gamma_\text{{\tiny GI}}}} \;=\; \frac{1}{\nu^2}\,.
\end{equation}
The R\'enyi entropy of the state can be expressed as~\cite{kim2018renyi}
\begin{equation}
S_\alpha \;=\; \frac{\ln\Tr[\hat\rho_\text{{\tiny GI}}^\alpha]}{1-\alpha}\;=\;
\frac{\sum_k\ln F_\alpha(\nu_k)}{\alpha-1}\;=\;\frac{2\ln F_\alpha(\nu)}{\alpha-1},
\end{equation}
where
\begin{equation}
F_\alpha(\nu) \;=\; \Bigl( \frac{\nu+1}{2} \Bigr)^\alpha-\Bigl( \frac{\nu-1}{2} \Bigr)^\alpha\,.
\end{equation}
In the limit $\alpha\to 1$ the R\'enyi entropy tends to the usual von-Neumann entropy~\cite{serafini}
\begin{equation}
S\;=\;-\Tr[\hat\rho_\text{{\tiny GI}}\ln\hat\rho_\text{{\tiny GI}}] \;=\; \sum_k f(\nu_k) \;=\; 2 f(\nu),
\end{equation}
where
\begin{equation}
f(\nu)\;=\;\Bigl( \frac{\nu+1}{2} \Bigr)\ln\Bigl( \frac{\nu+1}{2} \Bigr)-\Bigl( \frac{\nu-1}{2} \Bigr)\ln\Bigl( \frac{\nu-1}{2} \Bigr)\,.
\label{eq:fdef}
\end{equation}
Another special case is the R\'enyi-2 entropy which turns out to be related to the purity by
\begin{equation}
S_2 \;=\; -\ln \Tr[\hat \rho_\text{{\tiny GI}}^2]\;=\;2\ln \nu\;=\;-\ln \mu.
\end{equation}
The reduced density matrices $\hat\rho^{A,B}_\text{{\tiny GI}}=\Tr_{B,A}[\hat\rho_\text{{\tiny GI}}]$ of the two modes correspond to the reduced covariance matrices
\begin{equation}
\gamma^{A}_\text{{\tiny GI}} = \gamma^{B}_\text{{\tiny GI}} = 
\begin{pmatrix} \cosh(2r) &  \\ & \cosh(2r) \end{pmatrix}
\end{equation}
independent of the control parameter $p$. Hence, without classical communication between the modes, this parameter has no influence on local measurements. The corresponding R\'enyi entropy of the reduced states reads
\begin{equation}
S^{A}_\alpha=S^{B}_\alpha=\frac{\ln\bigl( \cosh^{2\alpha}(r)-\sinh^{2\alpha}(r) \bigr)}{\alpha-1},
\end{equation}
reducing to the von-Neumann entropy for $\alpha\to 1$, i.e.
\begin{equation}
S^{A}=S^{B}=2\cosh^2(r)\ln\cosh(r) - 2\sinh^2(r)\ln\sinh(r).
\end{equation}
%

\section{Entanglement criteria}
\label{sec:criteria}
%
\subsection{PPT criterion}
In this Section, we investigate the correlation properties of the state defined in Eqs.~(\ref{eq:gamma_GI2})-(\ref{densitymatrixGI}). First, we apply the Peres–Horodecki criterion~\cite{peres} which is also known as positive partial transpose (PPT) criterion. This criterion has been reformulated for CV systems in Ref.~\cite{simon2} and proven to be necessary and sufficient for all bipartite $1\times N$-mode Gaussian states in Refs.~\cite{werner2,lami}. For states in standard form, the PPT criterion yields a very simple expression which in the case of Eq.~(\ref{eq:gamma_GI2}) reads
\begin{align}
\begin{split}
	&\left(\cosh^2(2r)-p^2\sinh^2(2r)\right)^2\\
	&\phantom{oo}-2\left(\cosh^2(2r)+p^2\sinh^2(2r)\right)+1\,\geq\,0.
\end{split}
\label{eq:PPT}
\end{align}
If this inequality holds for a given set of parameters $(p,r)$, the criterion tells us that the resulting state $\hat\rho_\text{{\tiny GI}}$ is separable, otherwise it is entangled. This inequality can be rewritten as
\begin{equation}
	\cosh(2r)-p\sinh(2r)\,\geq\,1\quad\Leftrightarrow\quad p\,\leq\,\tanh(r),
\label{eq:ppt_sympl}
\end{equation}
where the l.h.s. is the smallest symplectic eigenvalue of the partially transposed covariance matrix $\tilde{\gamma}_\text{{\tiny GI}}$~\cite{adesso}. Note that this result for the entanglement threshold is already obtained as a by-product in Ref.~\cite{mista} when they consider the squeezing behavior of the non-Gaussian CV isotropic state.

\begin{figure}
	\begin{center}
		\includegraphics[width=86mm]{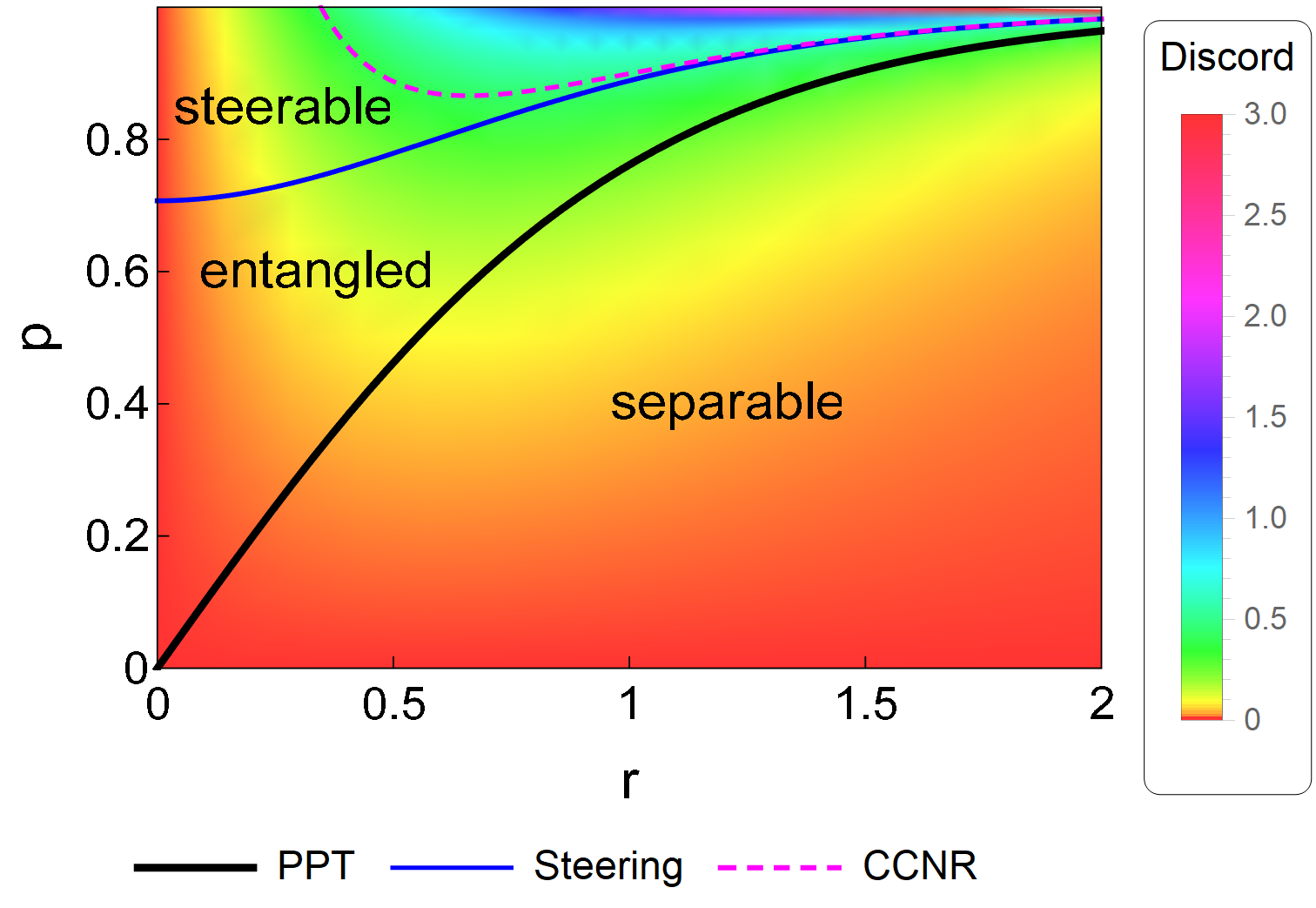}
		\caption{Entanglement criteria in the parameter space. In addition, quantum discord (see Sect.~\ref{subsec:qd}) is shown as a heatmap in order to stress the appearance of non-classical correlations in the separable domain.}
		\label{pic:ppt}
	\end{center}
\end{figure}

The separable domain with its boundary is shown in Fig.~\ref{pic:ppt}. We limited the range of the squeezing parameter $r$ to [0,2] since experimentally reachable squeezing values are currently limited to $r\approx 1.7$, see~\cite{anno,Vahlbruch_2016,shi,juan}. Obviously, as $r$ increases, the larger $p$ must be to obtain an entangled state. We will come back to this observation in the following section. In the limit $r\,\to\,\infty$, the r.h.s. of Eq.~(\ref{eq:ppt_sympl}) equals 1, meaning that the state becomes separable for all $p$.

We note that the inequality~(\ref{eq:ppt_sympl}) has been derived before in the literature in different forms. For example, for an initially pure TMS subjected to decoherence by Gaussian noise one can specify a temporal threshold from where on the entanglement vanishes~\cite{duan,serafini2005,isar2011,marian2015}.

\subsection{Steerability criterion}

Quantum steering in a bipartite system $\hat{\rho}_{AB}$ describes the ability of subsystem $A$ to influence the quantum state of subsystem $B$ by performing local measurements exclusively  on its own subsystem (see \cite{uola2020quantum} for a review). Clearly, steerability requires that the joint quantum state is entangled. In the case of Gaussian states, if one restricts the allowed measurements to Gaussian measurements, it has been shown in~\cite{wiseman2007steering,kogias2015quantification} that a state is steerable from $A$ to $B$ if and only if
\begin{equation}
\gamma + i\bigl(0_A \oplus \Omega_B \bigr) \; \geq \; 0
\end{equation}
is positive definite. Here, $0_A$ is a zero matrix in the subspace of $A$ and $\Omega_B$ is the symplectic matrix for subsystem~$B$. In the case of $\gamma_\text{{\tiny GI}}$, we are then led to the symmetric steering condition
\begin{equation}
p \;>\; \frac{1}{\sqrt{1+\frac{1}{\cosh(2r)}}}\,.
\end{equation}
The corresponding region is shown in Fig.~\ref{pic:ppt}. As expected, it is a subset of the entangled region. A related study of the steerability of a two-mode Gaussian state in the presence of thermal noise has been carried out recently in~\cite{abbasnezhad2019}.

\subsection{Realignment criterion}
%
For completeness, let us compare the PPT criterion with the computable cross norm or realignment (CCNR) criterion~\cite{chen2002matrix,rudolph2005further}. The realignment criterion states that for any separable state the trace norm of the realigned density matrix obeys the inequality $||\rho^R|| \leq 1$. Conversely, $||\rho^R|| > 1$ implies that the state is entangled. Zhang et al.~\cite{zhang2013detecting} reformulated the realignment criterion for Gaussian states, showing that for two-mode Gaussian states with a CM in the standard form~(\ref{eq:standard_form}) we have
\begin{equation}
||\rho^R|| \;=\; \frac{1}{2\sqrt{\bigl(\sqrt{ab}-|c_1| \bigr)\bigl( \sqrt{ab}-|c_2|  \bigr)}},
\end{equation}
so that
\begin{equation}
||\hat\rho_\text{{\tiny GI}}^R|| \;=\; \frac{1}{2\bigl(\cosh(2r)-p\sinh(2r)\bigr)}\,.
\end{equation}
Consequently, this criterion tells us that the inequality
\begin{equation}
p > \frac14 \coth(r) + 3\tanh(r)
\end{equation}
implies that the state is entangled. As can be seen in Fig.~\ref{pic:ppt}, the CCNR is compatible with the PPT criterion. However, whereas the PPT criterion is necessary and sufficient for all two-mode Gaussian states, the CCNR criterion is only necessary for separability in this case and thus, only detects a small subregion of the entangled domain.

\section{Measures of non-classical correlations}
\label{sec:measures}
%
\subsection{Entanglement of formation}

To quantify the \textit{amount} of entanglement contained in $\hat{\rho}_\text{{\tiny GI}}$, we compute the EOF defined by~\cite{bennet}
\begin{equation}
		E_F(\hat{\rho})\,=\,\underset{\qty{p_i,\ket{\psi_i}}}{\min}\qty{\sum\limits_ip_iS\qty(\dyad{\psi_i})\,\Big\vert\,\hat{\rho}\,=\,\sum\limits
		_ip_i\dyad{\psi_i}},
	\label{eq:EOF}
\end{equation}
where $S\qty(\dyad{\psi_i})$ is the local von-Neumann entropy of the pure-state components of the statistical ensemble yielding $\hat{\rho}$. The minimization has been explicitly solved for all symmetric two-mode Gaussian states in Ref.~\cite{giedke}. In this case, the EOF is given by the analytical expression 
\begin{equation}
		E_F(\hat{\rho})\,=\,\frac{(1+x)^2}{4x}\ln\left[\frac{(1+x)^2}{4x}\right]-\frac{(1-x)^2}{4x}\ln\left[\frac{(1-x)^2}{4x}\right],
	\label{eq:EOF_Gaussian}
\end{equation}
with $x\,=\,\min[1,\tilde{\nu}]$, where $\tilde{\nu}$ is the smallest symplectic eigenvalue of the partially transposed CM. For $\hat{\rho}_\text{{\tiny GI}}$, we have $\tilde{\nu}\,=\,\cosh(2r)-p\sinh(2r)$. Note that the PPT criterion~(\ref{eq:ppt_sympl}) is compatible with this measure. 

\begin{figure}
	\begin{center}
		\includegraphics[width=86mm]{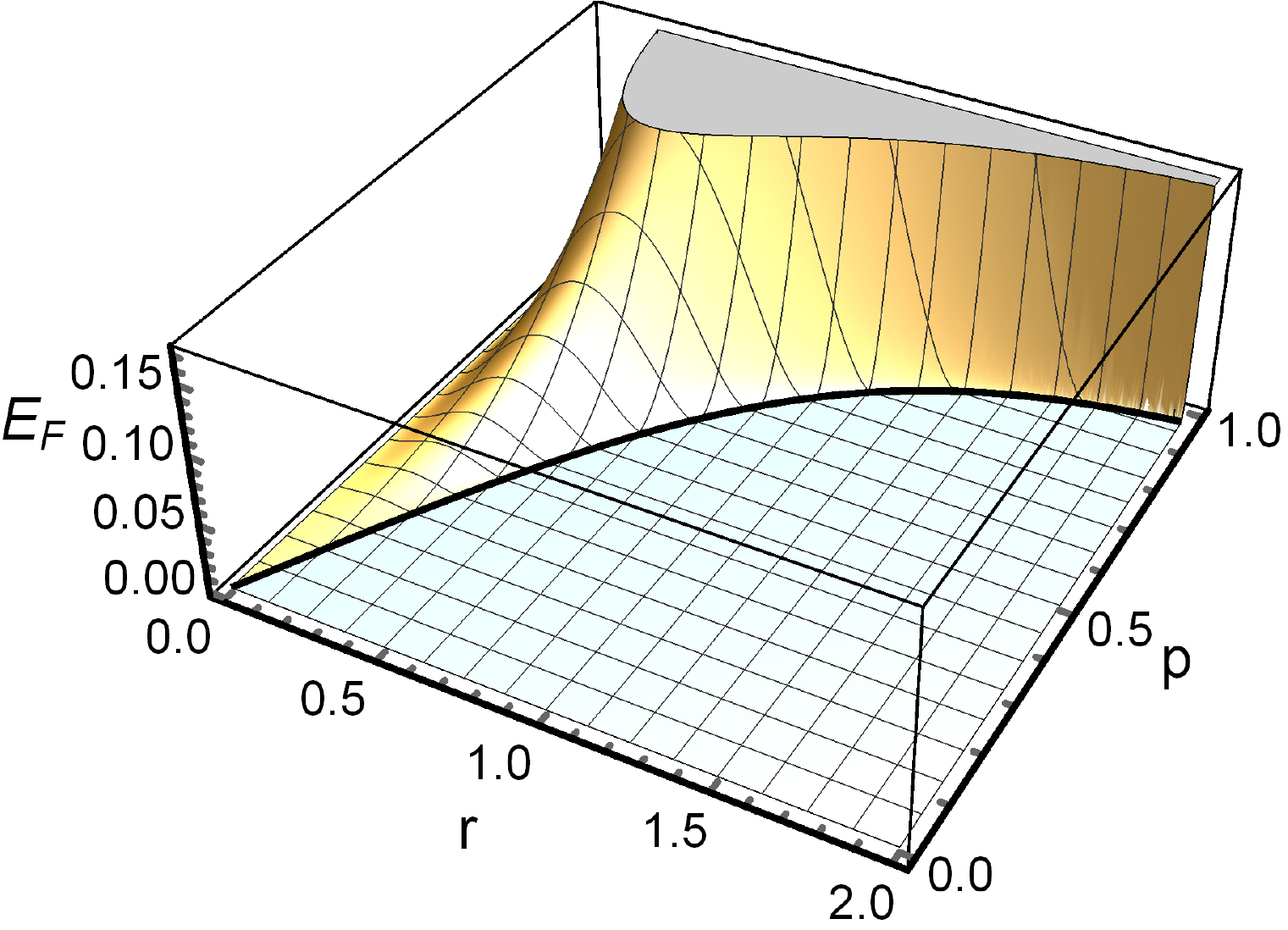}
		\caption{Entanglement of formation as function of $r$ and $p$. The black solid line marks the boundary between separable and entangled states which is given by $p=\tanh r$.}
		\label{pic:eof}
	\end{center}
\end{figure}

Fig.~\ref{pic:eof} shows $E_F(\hat{\rho}_\text{{\tiny GI}})$ as a function of the parameters $r$ and~$p$, where the latter controls the relative weight of the entangled component and the thermal noise in Eq.~(\ref{eq:gamma_GI1}). The black solid line marks the boundary between separable and entangled states according to Eq.~(\ref{eq:ppt_sympl}).

In the figures, we clearly see the competitive influence of the two parameters: the larger the squeezing $r$ is, the larger the probability  $p$ must be to obtain an entangled state. At first glance this behavior is counterintuitive since for stronger squeezing the TMS component in our state becomes more entangled and thus we would expect that more noise is needed to destroy the entanglement, i.e., the values of $p$ should decrease along the boundary with increasing $r$. Contrarily, the figure suggests that the entanglement in a weakly entangled TMS state was more robust against added noise as in a strongly entangled one. This apparent contradiction is resolved by observing that also the intensity of added thermal noise increases with $r$, and that this increase dominates the parameter dependence.

\subsection{Quantum discord}
\label{subsec:qd}
%
Next, we want to investigate $\hat{\rho}_{\text{\tiny GI}}$ with respect to possible quantum correlations \textit{beyond} entanglement. To this end, we analyze the quantum discord (QD)~\cite{ollivier,henderson}, a measure that is known to detect correlations that are due to quantum physical effects even in the absence of quantum entanglement. While the EOF reflects the average correlations of pure-state components in the statistical ensemble, minimizing over all possible compositions of the ensemble, the QD is defined as the total correlations given by the quantum mutual information minus the classical correlations extractable via measurement, maximizing over all possible quantum measurements:
\begin{equation}
D(\hat\rho)\;=\;S^B-S^{AB}+\inf_{\{\Pi_i\}}H_{\{\Pi_i\}}(A|B).
\end{equation}
The last term denotes the average conditional entropy of~$A$ after a generalized measurement $\{\Pi_i\}$ has been performed on $B$. 

A Gaussian version of the QD for all two-mode Gaussian states has been introduced in Ref.~\cite{datta,Giorda}. Here, the authors imposed the restriction that the aforementioned maximization is carried out over Gaussian measurements only, yielding a closed formula in terms of the entries of the CM in Eq.~(\ref{eq:standard_form}). In the case of $\gamma_{\text{\tiny GI}}$, this so-called Gaussian quantum discord is given by
\begin{align}   
\begin{split}
&D_G(\hat{\rho}_\text{{\tiny GI}})\,=\,f\bigl(\cosh 2r\bigr)+f\bigl(p^2-(p^2-1)\cosh 2r\bigr)\\
&\phantom{D_G\,=\,f}-2f\Bigl(\sqrt{\cosh^2(2r)-p^2\sinh^2(2r)}\Bigr),
\end{split}
\label{eq:qd}
\end{align}
where $f(\nu)$ is defined in Eq.~(\ref{eq:fdef}). 

While the EOF vanishes in an extended region of the parameter space, the QD is non-zero everywhere except for the boundaries $r=0$ and $p=0$, see Fig.~\ref{pic:ppt}. This is reasonable because for $r=0$ the CM $\gamma_{\text{\tiny GI}}$ is simply the product of two coherent states while for $p=0$ it is the product of two thermal states. In fact, the QD is known to vanish only on product states, as has been proven in Ref.~\cite{datta} independently of the restriction to Gaussian measurements. 

\begin{figure}
\centering
\includegraphics[width=86mm]{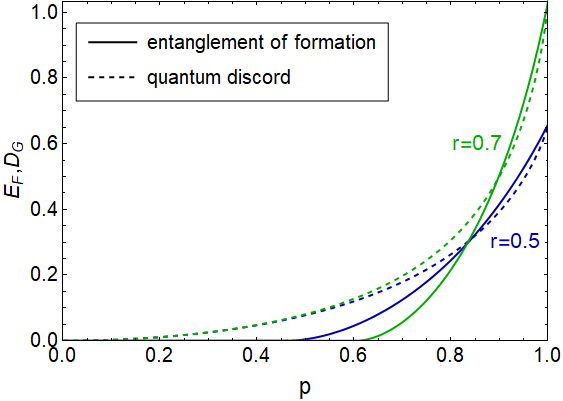}
\caption{Comparison of the entanglement of formation and the Gaussian quantum discord for fixed values of $r$. The qualitative behavior is the same as the one of a two-qubit isotropic state~\cite{ollivier}.}
\label{pic:fix_r}
\end{figure}

The behavior of the QD for constant squeezing parameter $r$ is shown in Fig.~\ref{pic:fix_r}. As can be seen, the Gaussian QD increases with $p$. In contrast to the EOF, the QD is non-zero in the separable region, responding to the quantumness of the correlations. Thus, for a fixed value of $r$, the qualitative behavior of EOF and QD is the same as in a two-qubit isotropic state~\cite{ollivier}.

Keeping instead $p$ fixed and varying $r$, the Gaussian QD first rises, see Fig.~\ref{pic:ppt}. For large values of $r$, however, it behaves similar to the EOF in the sense that it becomes negligibly small except for probabilities close to one. For experimentally achievable squeezing values, however, we can always chose the probability parameter such that the EOF is zero while QD has a considerable finite value. Therefore, in contrast to the non-Gaussian CV isotropic state which is is entangled for all $p>0$, the Gaussian version proposed here can serve as an interesting test bed to study non-classical correlations in separable states within the Gaussian setting. 

\begin{figure}
\centering
\includegraphics[width=86mm]{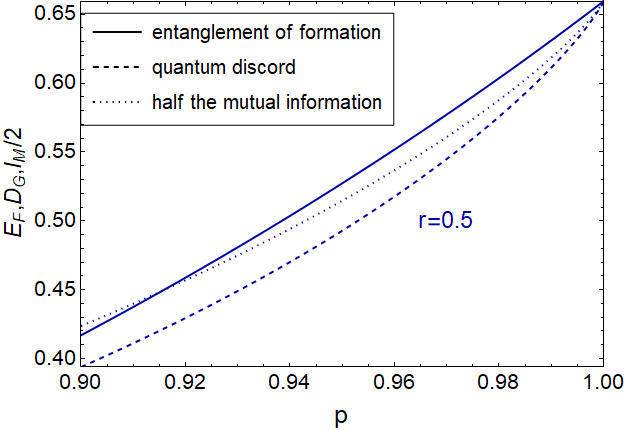}
\caption{The entanglement of formation exceeds the Gaussian quantum discord and half the mutual information in a certain parameter range.}
\label{pic:fix_r_MI}
\end{figure}

Another analogy between $\hat{\rho}_{\text{\tiny GI}}$ and the finite-dimensional isotropic state shows up when comparing EOF and mutual information which for Gaussian states can be calculated using the expression for the von-Neumann entropy~\cite{serafini}. In Fig.~\ref{pic:fix_r_MI}, we demonstrate that for certain values of the parameters the EOF exceeds half the mutual information $I_M/2$. This is probably an artifact of the EOF if we accept the mutual information as a proper measure of total correlations in a quantum state, as has been discussed in detail for the finite-dimensional isotropic and Werner states in Ref.~\cite{li}.

\section{Gaussian channel}
\label{sec:gaussian_channel}
In Ref.~\cite{giedke_cirac}, the class of all physical Gaussian operations has been characterized with the help of the isomorphism between completely positive maps and bipartite states~\cite{khatri_wilde}. According to Ref.~\cite{giedke_cirac}, a map $\mathcal{G}$ is a \textit{Gaussian} completely positive map if the bipartite state $\hat{\rho}$ isomorphic to $\mathcal{G}$ is determined by a proper CM $\Gamma$ (fulfilling the physicality condition~(\ref{eq:phys_cond})). Conversely, to each such Gaussian state $\hat{\rho}$ corresponds a Gaussian completely positive map $\mathcal{G}$. Such a map acts on the CM $\gamma$ of a Gaussian input state as follows\,\cite{giedke_cirac}:
\begin{align}
    \gamma\,\to\,&\gamma'\,=\,\tilde{\Gamma}_{11}-\tilde{\Gamma}_{12}\qty(\tilde{\Gamma}_{22}+\gamma)^{-1}\tilde{\Gamma}_{12}^T,
    \label{eq:gcp_map}\\
    &\text{where}\,\,\tilde{\Gamma}\,=\,\qty(\begin{array}{cc}
       \tilde{\Gamma}_{11}  & \tilde{\Gamma}_{12} \\
        \tilde{\Gamma}_{12}^T & \tilde{\Gamma}_{22}
    \end{array})\nonumber
\end{align}
is the partial transpose of the CM describing the Gaussian state isomorphic to the map $\mathcal{G}$. Note that $\Gamma$ is twice as large as $\gamma$ which characterizes the input state of the Gaussian channel described by $\mathcal{G}$.

Obviously, such a relation also holds for $\hat{\rho}_\text{{\tiny GI}}$ ~\cite{mark_wilde} which is isomorphic to the map $\mathcal{G}_\text{{\tiny GI}}$ that acts on single-mode Gaussian systems. Computing the partial transpose of $\gamma_\text{{\tiny GI}}$, Eq.~(\ref{eq:gamma_GI2}), and applying Eq.~(\ref{eq:gcp_map}), we directly obtain the output mode of the Gaussian channel described by $\mathcal{G}_\text{{\tiny GI}}$. Taking as input a coherent state with CM $\sigma_\text{{\tiny coh}}\,=\,\mathbb{1}_2$, the output reads
\begin{equation}
        \sigma_\text{{\tiny coh}}'\,=\,p^2\sigma_\text{{\tiny coh}}+(1-p^2)\cosh(2r)\mathbb{1}_2.
    \label{eq:output_coh_mode}
\end{equation}
This is a convex combination of the original CM and the CM of a thermal mode. We therefore see another analogy to the finite-dimensional isotropic state which is known to be isomorphic to the depolarizing channel~\cite{horodeckis2,hashagen}. This channel maps a state $\hat{\rho}$ onto a convex combination of itself and the maximally mixed state. However, we emphasize that the expression in~Eq.\,(\ref{eq:output_coh_mode}) is not a convex combination of two density operators but of two CMs and the second term is a \textit{partially} mixed state, where the degree of mixedness depends on the squeezing parameter $r$. 

If the input is not a coherent but a thermal mode, the output does not display the clear superposition of original and thermal CM. Rather, one can then observe that -- depending on $r$ and $p$ -- the output mode can be \textit{less} noisy than the input mode. We assume that the action of the channel described by $\mathcal{G}_\text{{\tiny GI}}$ -- adding or blocking of classical noise in dependence of $r$ and $p$ -- is related to the correlation structure of the state $\hat{\rho}_\text{{\tiny GI}}$ isomorphic to $\mathcal{G}_\text{{\tiny GI}}$. However, figuring out the exact relation requires further investigation.

\section{Discussion}

In this paper, we have analyzed a Gaussian version of an isotropic state which is controlled by a mixing parameter $0\leq p \leq 1$ and a squeezing parameter $r \geq 0$. Unlike the non-Gaussian two-parametric isotropic state introduced by Mi\v{s}ta \textit{et al.}, the Gaussian version studied here features a finite threshold $p_c=\tanh r$ where entanglement sets in. Below this threshold the state is separable but it still exhibits quantum correlations, as can be detected by the quantum discord. It can therefore serve as a simple bipartite test state to study non-classical correlations in the absecnce of entanglement within the Gaussian setting. 

To understand the quantum nature in the separable region $0 < p \leq p_c$, it is instructive to compare the situation with the two-qubit isotropic state (cf. Eq.~(\ref{eq:iso_state}))
\begin{equation}
\hat{\rho}_\text{{\tiny I}} \;=\; p\dyad + \,+\, (1-p)\frac{\mathbb{1}_4}{4}
\end{equation}
with $\ket\pm \;=\; \frac{1}{\sqrt{2}}\bigl( \ket{00}\pm\ket{11} \bigr)$, where $p_c=\frac13$. 

Entanglement means that local operations on one of the subsystems influence the other subsystem. In the fully entangled case $p=1$, where $\hat{\rho}_\text{{\tiny I}} =\dyad +$ is just a Bell state, any local operation $C$ on one side can be swapped to the other side by means of
\begin{equation}
(C \otimes \mathbb{1}_2)\ket + \;=\; (\mathbb{1}_2  \otimes C^T)\ket + \,.
\end{equation}
Decreasing $p$, this simple shift is replaced by the possibility of quantum steering, describing the control of part $B$ by local operations in part $A$ and vice versa, which is possible in the range $\frac12 < p \leq 1$. Decreasing $p$ further there is a non-steerable but entangled region $\frac13 < p \leq \frac12$. In this range one-sided steering is no longer possible, but quantum correlations can still be detected by two-sided local operations assisted by classical communication. Finally, for $p\leq \frac13$ the state becomes separable, i.e., it can be written in the form
\begin{equation}
\hat{\rho}_\text{{\tiny I}} \;=\; \sum_{i} p_{i} \dyad{\psi^A_{i}} \otimes \dyad {\psi^B_i}
\end{equation}
with probabilities $0\leq p_{i} \leq 1$ normalized by $\sum_{i}p_{i}=1$. If such a representation is found, it does provide a practical description how to generate the quantum state exclusively by local operations and classical communication, hence it is clear that it cannot be used for mutual quantum control. Nevertheless, such a state can be non-classical in the sense that the $\ket{\psi^A_i}$ and $\ket{\psi^B_i}$ are not necessarily orthogonal in the respective subsystems. This is also reflected in the eigenvalue decomposition
\begin{equation}
\begin{split}
\hat{\rho}_\text{{\tiny I}} 
\;=\;&
\frac{1+3p}{4} \dyad{+} + \frac{1-p}{4}\dyad - \\ & + \frac{1-p}{4}\Bigl( \dyad{01}+\dyad{10} \Bigr)\,.
\end{split}
\end{equation}
As can be seen, the decomposition involves the fully entangled Bell states $\ket\pm$, but for small $p$ they increasingly compensate one another so that quantum communication of both sides is no longer possible. However, the non-classicality is still detectable by means of the quantum discord.

The Gaussian CV isotropic state considered in this paper exhibits completely analogous phenomenology (see~Fig.~\ref{pic:ppt}~and~Fig.~\ref{pic:fix_r}) and provides an example where the onset of entanglement can be studied by similar means. In this context, it would be interesting to find the separable representation in the regime $p\leq\tanh r$, as outlined in Ref.~\cite{guhne2007covariance}.

Regarding the fact that the output state of the teleportation protocol adds Gaussian noise to a pure input state and that the Gaussian CV isotropic state can be seen as pure two-mode squeezed state with additional thermal noise, it would be interesting to ask if one can obtain this kind of state when both modes of a two-mode squeezed state are taken as input to the teleportation process.

Moreover, a more detailed study of the Gaussian channel isomorphic to the Gaussian CV isotropic state -- exhibiting another analogy to the finite-dimensional isotropic state when a coherent state is considered as input -- could lead to further insights into the effects of the quantum correlations contained in this state. 

\vspace{4mm}

\textbf{Acknowledgments:} We thank M. M. Wilde for pointing out that it is possible to obtain the Gaussian isotropic state from many copies of the non-Gaussian previously studied state and further for making us aware of the possibility to study the corresponding channel according to the Choi-Jamio\l kowski isomorphism.

\appendix
\section{Calculation of the density matrix}
\label{sec:density}
%
In this Appendix, we outline how the density matrix~$\hat\rho_\text{{\tiny GI}}$ in Eq.~(\ref{densitymatrixGI}) can be computed explicitly. To this end, we first determine the matrix elements $\langle\mu\nu|\hat\rho_\text{{\tiny GI}}|\kappa\tau\rangle$ in the coherent representation
\begin{equation}
|\mu\nu\rangle \;=\; \bigl[ \hat D(\mu)\otimes \hat D(\nu) \bigr]\,|0,0\rangle\,,
\end{equation}
where $|0,0\rangle$ is the two-mode vacuum state and $\hat D(\alpha)=e^{\alpha\hat a^\dagger -\alpha^*\hat a}$ is the displacement operator for a single mode. With $\alpha=\frac{1}{\sqrt{2}}(x_1+ip_1)$ and $\beta=\frac{1}{\sqrt{2}}(x_2+ip_2)$ the density matrix~(\ref{densitymatrixGI}) can be written as
\begin{equation}
\hat\rho_\text{{\tiny GI}} \;=\; \frac{1}{\pi^2} \int d^2\alpha \, \int d^2\beta \, \chi(\alpha,\beta) \, \hat D^\dagger(\alpha,\beta),
\end{equation}
where we used the usual notations $\hat D(\alpha,\beta)=\hat D(\alpha)\otimes \hat D(\beta)$ and $\int d^2\alpha=\frac12 \int d x\int d p$. The desired matrix elements are then given by
\begin{align}
\label{app:eq3}
\langle\mu\nu|\hat\rho_\text{{\tiny GI}}|\kappa\tau\rangle &=
\langle 0,0|\hat D^\dagger(\mu,\nu)\hat\rho_\text{{\tiny GI}}\hat D(\kappa,\tau)|0,0\rangle\\
&=\frac{1}{\pi^2} \int d^2\alpha \, \int d^2\beta \, \chi(\alpha,\beta) \, \times\nonumber\\
&\hspace{10mm} \langle 0 | \hat D^\dagger(\mu) \hat D^\dagger(\alpha) \hat D(\kappa) |0\rangle \times \nonumber \\
&\hspace{10mm}\langle 0 | \hat D^\dagger(\nu) \hat D^\dagger(\beta) \hat D(\tau) |0\rangle \,. \nonumber
\end{align}
Inserting the CM (\ref{eq:gamma_GI2}) into (\ref{eq:CF}) and setting $\vec d=0$, the characteristic function reads
\begin{equation}
\begin{split}
\chi(\alpha,\beta)\;&=\;
\exp\Bigl[ -\frac{1}{2}\Bigl(\alpha\alpha^*+\beta\beta^* \Bigr)\cosh(2r) \\
& \hspace{12mm}+ \frac{p}2 \Bigl( \alpha\beta+\alpha^*\beta^* \Bigr)\sinh(2r) \Bigr]\,.
\end{split}
\end{equation}
If we insert this expression into (\ref{app:eq3}), and if we apply the relations $\hat D^\dagger(\alpha)=\hat D(-\alpha)$ and
\begin{equation*}
\langle 0| \hat D(\alpha) \hat D(\beta) \hat D(\gamma) | 0\rangle = e^{-\frac12(\alpha\alpha^*+\beta\beta^*+\gamma\gamma^*) - \alpha^*\beta - \alpha^*\gamma -  \beta^*\gamma }
\end{equation*}\\
we get the expression
\begin{equation}
\begin{split}
\bra{\mu\nu}\hat\rho_\text{{\tiny GI}}\ket{\kappa\tau}  &=
 e^{-\frac12(\mu\mu^*+\kappa\kappa^*+\nu\nu^*+\tau\tau^*) + \mu^*\kappa+\nu^*\tau} \\ & \times
 \frac 1 \pi \int d^2\beta\,e^{ -\frac12 \beta\beta^* [\cosh(2r)+1] - \nu^*\beta + \beta^*\tau } \\ & \times
 \frac 1 \pi \int d^2\alpha\, e^{-\frac12 \alpha\alpha^*[ \cosh(2r)+1]- \mu^*\alpha +\alpha^*\kappa } \times \\ & \hspace{19mm}e^{+\frac{p}{2}(\alpha\beta+\alpha^*\beta^*)\sinh(2r)  }\,.
\end{split}
\end{equation}
The second integral over $\alpha$ is of Gaussian type and can be evaluated using the formula
\begin{equation}
\label{gaussianintegral}
\int d^2 \alpha \, e^{-q \, \alpha\alpha^* + \alpha\beta^* - \alpha^*\gamma } \;=\;
\frac{\pi}{q}\exp\Bigl( -\frac{\gamma\beta^*}{q} \Bigr)\,.
\end{equation}
Inserting the result back into the remaining expression gives again a Gaussian integral over $\beta$ which can be evaluated by similar means. After some algebra, we arrive at the explicit expression

\begin{widetext}
\begin{equation}
\begin{split}
\bra{\mu\nu}\hat\rho_\text{{\tiny GI}}\ket{\kappa\tau} &=
 \,\, \frac{\exp\Bigl[-\frac12(\mu\mu^*+\kappa\kappa^*+\nu\nu^*+\tau\tau^*) + \mu^*\kappa+\nu^*\tau- \frac{\kappa\mu^*}{\cosh^2(r)}-
 \,\, \ \frac{(\nu^*-p\kappa\tanh(r))(\tau-p\mu^*\tanh(r) )}{\cosh^2(r)-p^2\sinh^2(r)} \Bigr]}
 {\cosh^2(r)\Bigl( \cosh^2(r)-p^2\sinh^2(r) \Bigr)}.
\end{split}
\end{equation}
For the special case $p=0$ this expression reduces to
\begin{equation}
\bra{\mu\nu}\hat\rho_\text{{\tiny GI}}\ket{\kappa\tau} \;=\; \frac{1}{\cosh^4(r)}
\exp\Bigl[ -\frac12(\mu\mu^*+\kappa\kappa^*+\nu\nu^*+\tau\tau^*)+(\kappa\mu^*+\tau\nu^*)\tanh^2(r) \Bigr],
\end{equation}
while for $p=1$ we get
\begin{equation}
\bra{\mu\nu}\hat\rho_\text{{\tiny GI}}\ket{\kappa\tau} \;=\; \frac{1}{\cosh^2(r)}
\exp\Bigl[ -\frac12(\mu\mu^*+\kappa\kappa^*+\nu\nu^*+\tau\tau^*) +(\mu^*\nu^*+\kappa\tau)\tanh(r)\Bigr]\,.
\end{equation}
\end{widetext}

For these special cases it is straightforward to arrive at the more common representation in the Fock basis. For $p=0$ the state factorizes into
\begin{equation}
\hat\rho_\text{{\tiny GI}}\,=\,\hat\rho_0 \otimes \hat\rho_0
\end{equation}
where 
\begin{equation}
\begin{split}
\hat\rho_0 &= \frac{1}{\cosh^2(r)\pi^2}\int d^2\mu\int   	
 d^2\kappa\,\dyad{\mu}{\kappa} \\ & \hspace{4mm} \times \exp(-\frac{1}{2}(\mu\mu^*+\kappa\kappa^*)+\mu^*\kappa\tanh^2r)\,.
\end{split}
\end{equation}
In order to express $\hat\rho_\text{{\tiny GI}}$ in the Fock basis, we have to perform a suitable basis transformation. For $p=0$ this transformation can be carried out separately in each tensor slot. Inserting $\ket{\alpha}\,=\,e^{-\frac{1}{2}\alpha\alpha^*}\sum\limits_{n=0}^{\infty}\frac{\alpha^n}{\sqrt{n!}}\ket{n}$, we get
\begin{align}
\hat\rho_0 &= \frac{1}{\cosh^2(r)\pi^2}\sum\limits_{m,n=0}^{\infty}\frac{\dyad{m}{n}}{\sqrt{m!n!}}\int d^2\mu\int	 d^2\kappa\,\mu^m\kappa^{*n} \nonumber \\ 
 & \hspace{4mm} \times e^{-\frac{1}{2}(\mu\mu^*+\kappa\kappa^*)+\mu^*\kappa\tanh^2r} 
\end{align}
which can also be written using derivatives as
\begin{align}
 \hat\rho_0 &= \frac{1}{\cosh^2(r)\pi^2}\sum\limits_{m,n=0}^{\infty}\frac{\dyad{m}{n}}{\sqrt{m!n!}}
 \partial_\alpha^m \partial_\beta^n \int d^2\mu\int	 d^2\kappa \nonumber \\ 
 & \hspace{4mm} \times e^{-\frac{1}{2}(\mu\mu^*+\kappa\kappa^*)+\mu^*\kappa\tanh^2r+\mu\alpha+\kappa^*\beta}\Bigr|_{\alpha=\beta=0}.
\end{align}
Now, the two integrals are of Gaussian type and can be carried out using Eq.~(\ref{gaussianintegral}), giving $\pi^2e^{\alpha\beta\tanh^2r}$, hence
\begin{equation}
\begin{split}
\hat\rho_0&=\frac{1}{\cosh^2(r)}\sum\limits_{m,n=0}^{\infty}\frac{\dyad{m}{n}}{\sqrt{m!n!}}\\ & \hspace{4mm} \times \Bigl( \partial_\alpha^m  \left(\alpha\tanh^2(r)\right)^n e^{\alpha\beta\tanh^2r} \Bigr)\Bigr|_{\alpha=\beta=0}
\end{split}
\end{equation}
Carrying out the differentiation, one can see that only terms with $m=n$ contribute, leading us to the final result
\begin{equation}
\hat\rho_0 \;=\;\sum\limits_{n=0}^{\infty}\frac{\left(\sinh^2(r)\right)^n}{\left(\cosh^2(r)\right)^{n+1}}\dyad{n},
\end{equation}
which is a one-mode thermal state. As expected, for $p=0$ we have $\hat\rho_\text{{\tiny GI}}\,=\,\hat\rho_0\otimes \hat\rho_0\,=\,\hat\rho_\text{{\tiny TMT}}$. 

For  $p=1$ the calculation follows similar lines, the difference being that the entangled state does not factorize, instead we have to perform the basis transformation described above for the bra- and ket-vector of $\hat\rho_\text{{\tiny GI}}$, respectively. Using the same methods as outlined above, one can show that for $p=1$ we obtain
\begin{equation}
\hat\rho_\text{{\tiny GI}}\,=\,\left(1-\tanh^2(r)\right)\sum\limits_{m,n=0}^{\infty}\tanh^{m+n}(r)\dyad{m,m}{n,n}\,,
\end{equation}
that is, we get indeed $\hat\rho_\text{{\tiny GI}}=\hat\rho_\text{{\tiny TMS}}$.


\bibliography{Paper.bib}

\end{document}